\documentclass[12pt]{iopart}

\usepackage[]{graphics,graphicx,epsfig}            
\usepackage{float}
\restylefloat{figure}
\usepackage{graphicx}
\usepackage{caption}
\usepackage{subcaption}

\usepackage{graphicx}
\usepackage{epstopdf}

\begin{document}

\title{Lifetime of topological quantum memories in thermal environment }

\author{Abbas Al-Shimary, James R. Wootton and Jiannis K. Pachos}

\address{School of Physics and Astronomy, University of Leeds, Leeds LS2 9JT, U.K}
\ead{j.k.pachos@leeds.ac.uk}

\address{Department of Physics, University of Basel, Klingelbergstrasse 82, CH-4056 Basel, Switzerland}
\ead{james.wootton@unibas.ch}

\begin{abstract}
Here we investigate the effect lattice geometry has on the lifetime of two-dimensional topological quantum memories. Initially, we introduce various lattice patterns and show how the error-tolerance against bit-flips and phase-flips depends on the structure of the underlying lattice. Subsequently, we investigate the dependence of the lifetime of the quantum memory on the structure of the underlying lattice when it is subject to a finite temperature. Importantly, we provide a simple effective formula for the lifetime of the memory in terms of the average degree of the lattice. Finally, we propose optimal geometries for the Josephson junction implementation of topological quantum memories.
\end{abstract}

\maketitle

\section{Introduction}
Topologically ordered systems are promising candidates for quantum memories. The main idea of Kitaev in \cite{Kitaev} was to use spin (or qubit) Hamiltonians with topologically ordered ground-states to encode quantum information \cite{Pachos}.  Logical qubits encoded into the ground-state of such systems become virtually resilient against environmentally induced local static perturbations to their Hamiltonian. The simplest such memory, known as the toric code, is considered as the testbed for studying the properties of topological order against various models of errors. For concreteness, we shall restrict our study to this model.

In the presence of probabilistic errors, robustness against perturbations is no longer sufficient to ensure self-correction \cite{Dennis}. Assume independent probabilistic bit-flip (Pauli $X$ operator) and phase-flip (Pauli $Z$ operator) errors during the preparation of the initial state on the physical qubits of the toric code. Retrieval of encoded information is possible only if the relative number of  $X$ and $Z$ errors is below a certain threshold. This error-tolerance threshold has been calculated for the toric code on a square lattices, to be approximately $11\%$ \cite{Dennis}, as well as for various other regular lattices \cite{Fujii}.

Consider now a topological memory coupled to a thermal bath. If the system is initially in a pure state then the thermal environment will start introducing errors. After a certain time one expects that a sufficiently large number of errors will be accumulated causing the memory to fail \cite{Kay-Colbeck}. This time is known as the memory lifetime $\tau$ \cite{Chesi}.

Errors in topological systems, such as the toric code, take the form of topological defects known as anyons \cite{Pachos}. A finite temperature can cause anyons to be created, propagated or annihilated. A fundamental flaw of two-dimensional topological memories that rules out self-correction under probabilistic errors is the lack of energy barriers that could suppress diffusion of anyons over large distances. For instance, consider a process that involves a creation of anyon pair from the ground state, a transport of one anyon along a non-contractible loop on the torus, and a final annihilation of the pair. This process enacts a non-trivial transformation on the subspace of ground states. However, it can be implemented by a stream of local errors at a constant energy cost \cite{Kay-Colbeck, Bravyi-Terhal}.

In fact, Alicki {\em et al.} \cite{Alicki} demonstrated that in one and two spatial dimensions, no stabiliser Hamiltonian can serve as a self-correcting quantum memory. They showed that the relaxation time towards the equilibrium state is a constant independent of the lattice size. The physical mechanism behind this instability is the absence of interactions between the anyonic thermal excitations. Indeed, \cite{Chesi,Hamma} demonstrated self-correction properties of two-dimensional topological systems by allowing long-range interactions between anyons.

The main goal of this paper is to investigate the possibility of hindering the diffusion of anyons by considering the toric code on general, non-regular lattices. By localising anyons to certain regions on the lattice, they are less likely to travel far apart, leading to better error correction properties. However, deforming the surface code breaks the symmetry between the $X$ and $Z$ error correction properties, creating an asymmetry in the error threshold values and their corresponding lifetimes \cite{Fujii}. This is because stabiliser operators are no longer symmetric under the lattice duality transformation (i.e. exchange of vertices and faces with each other). As an application, we demonstrate how such lattice geometries can be employed to enhance topological systems with intrinsically biased couplings due to physical implementation such as Josephson-junction arrays. 

\begin{figure}[t]
\centering
\includegraphics[width=8cm]{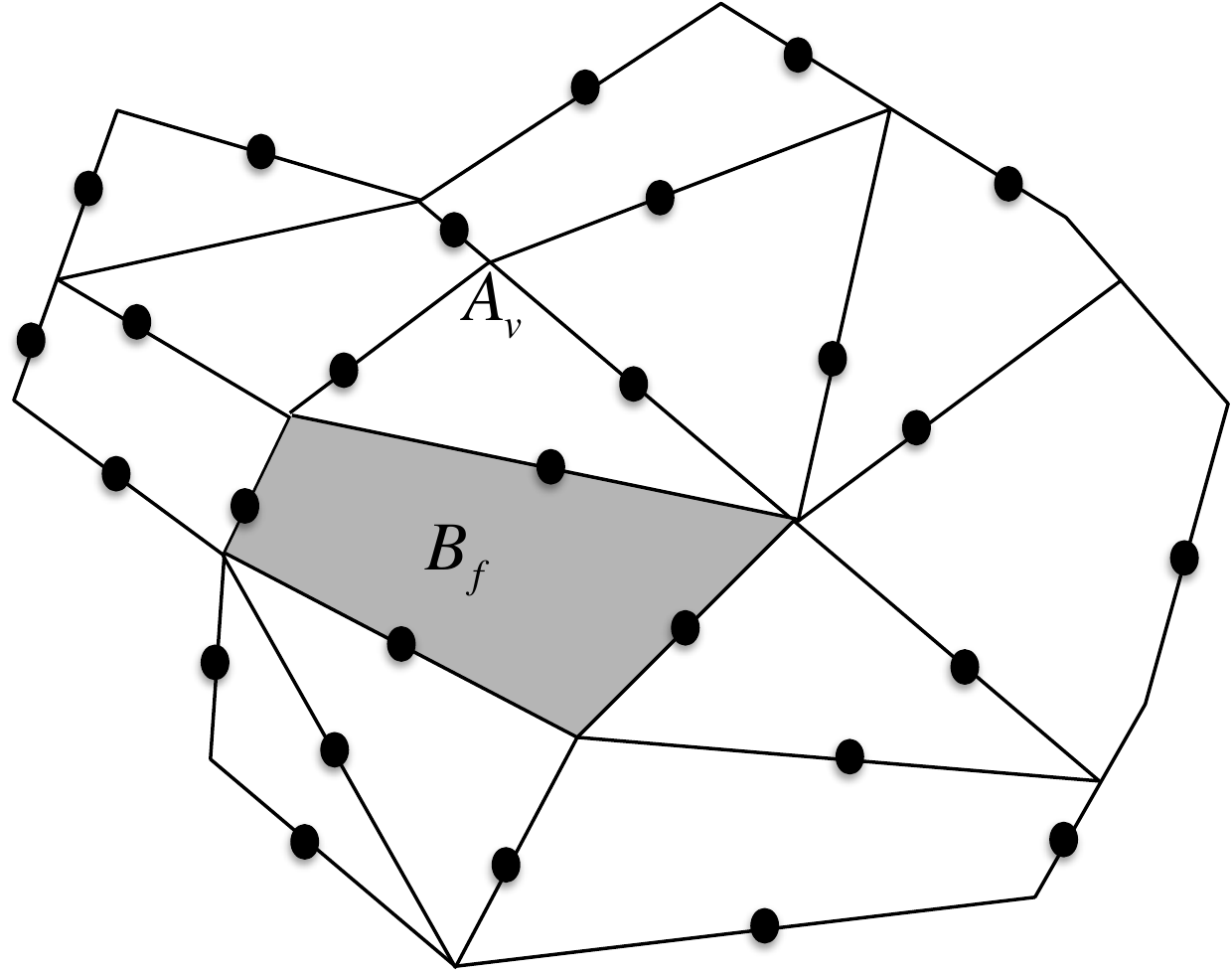}
\caption{\small The toric code defined for a general lattice structure. The plaquette, or face operator $B_f$ acts non-trivially on the qubits of the face $f$. The vertex operator $A_v$ acts non-trivially on the qubits that neighbour vertex $v$.}
\label{fig:GeneralLattice}
\end{figure}

\section{The two-dimensional toric code}

The starting point of our study is Kitaev's two-dimensional toric code \cite{Kitaev}. Consider a general lattice $\mathcal{L}(E,V,F)$, as shown in Fig. \ref{fig:GeneralLattice}, with periodic boundary conditions (a torus), where $E$, $V$, and $F$ are sets of edges, vertices, and faces, respectively. A qubit is associated with each edge. The stabiliser operators of the code on the lattice $\mathcal{L}$ are defined for each vertex $v \in V$ and face $f \in F$, respectively, as

\[ A^{\mathcal{L}}_{v} = \bigotimes_{i \in E_v} X_i, B^{\mathcal{L}}_{f} = \bigotimes_{j \in E_f} Z_i,\]
where $X_i$ and $Z_i$ denote Pauli operators on the $i$-th and $j$-th qubits respectively, and $E_v$ and $E_f$ indicate the sets of edges originating from vertex $v$ and surrounding the face $f$ respectively \cite{Fujii}. We define lattice $\mathcal{L}$ to be the primal lattice. Its dual lattice, denoted by $ \bar{\mathcal{L}}( \bar{E} , \bar{V} , \bar{F} )$, is defined by exchanging the vertices and faces of $\mathcal{L}$ with each other. The stabiliser space, where information is stored, is composed of states which satisfy $A^{\mathcal{L}}_{v} | \psi \rangle = | \psi \rangle$ and $B^{\mathcal{L}}_{f} | \psi \rangle = | \psi \rangle$ for all $v$ and $f$. Violations of these stabilizers on vertices or faces are associated with different types of anyons. The so called $e$ anyons reside on vertices and $m$ anyons on faces. Alternatively, the error syndromes of $e$ and $m$ anyons are associated with the vertices on the primal and dual lattices, respectively. Hence, the thermal delusion of the $e$ and $m$ anyons depends on the connectivity of the primal and dual lattices, respectively. Error correction corresponds to a pairwise annihilation of anyons which does not form topologically non-contractable loops. 

The toric code has been so far intensively investigated mainly on regular lattices \cite{Dennis,Fujii}. In the case of the square lattice, which is a regular self-dual lattice $\mathcal{L} = \bar{\mathcal{L}}$, assuming independent errors, the toric code can be mapped to a random bond Ising model, resulting in symmetric error tolerances to $e$ and $m$ anyons and a threshold value of $11\%$ \cite{Dennis}. When the toric code is coupled to a thermal bath, one observes a sharp transition in time known as the memory lifetime, $\tau$. For square lattice, the lifetime of the primal and dual lattices is same and for temperature $T/J= 0.3$ it is approximately equals $\tau \approx 5.8$ \cite{Chesi}. 

\begin{figure}
\centering
        \begin{subfigure}[b]{0.23\textwidth}
                \centering
                \includegraphics[width=\textwidth]{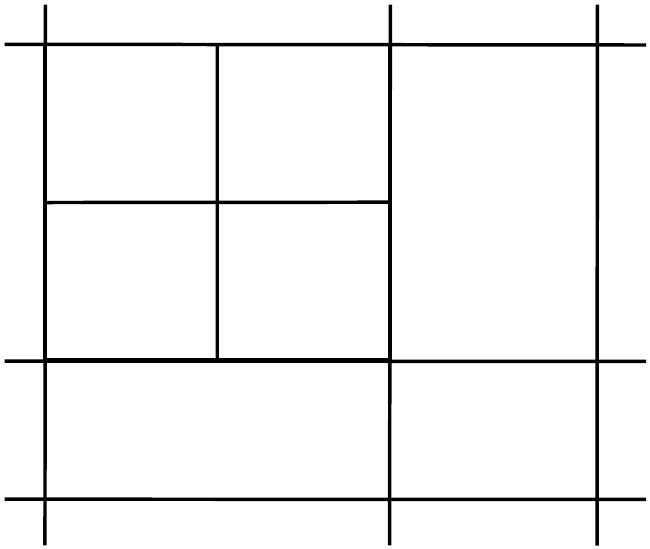}
                \caption{Reduced Square}
                \label{fig:ReducedSquare}
        \end{subfigure}
        ~ 
        \begin{subfigure}[b]{0.24\textwidth}
                \centering
                \includegraphics[width=\textwidth]{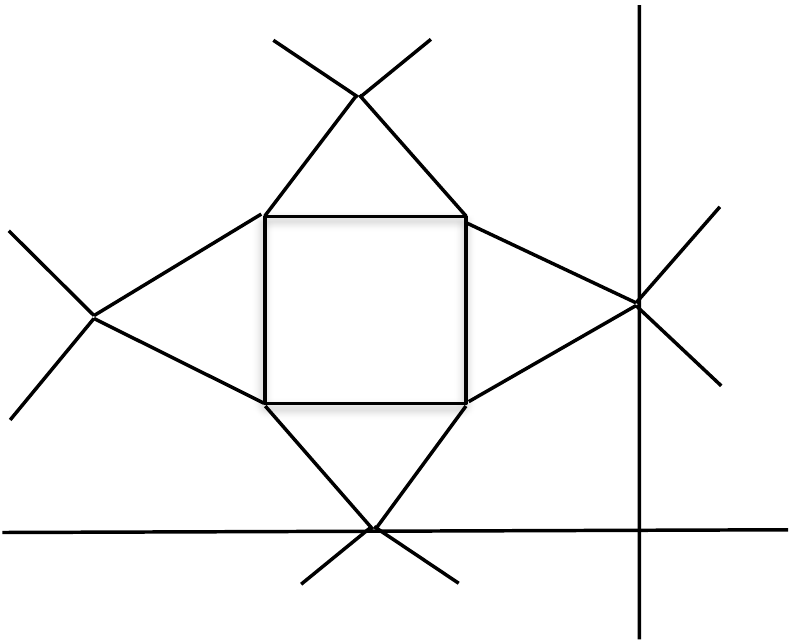}
                \caption{Dual Reduced Square}
                \label{fig:DualReducedSquare}
        \end{subfigure}\\
             ~ 
                \begin{subfigure}[b]{0.23\textwidth}
                \centering
                \includegraphics[width=\textwidth]{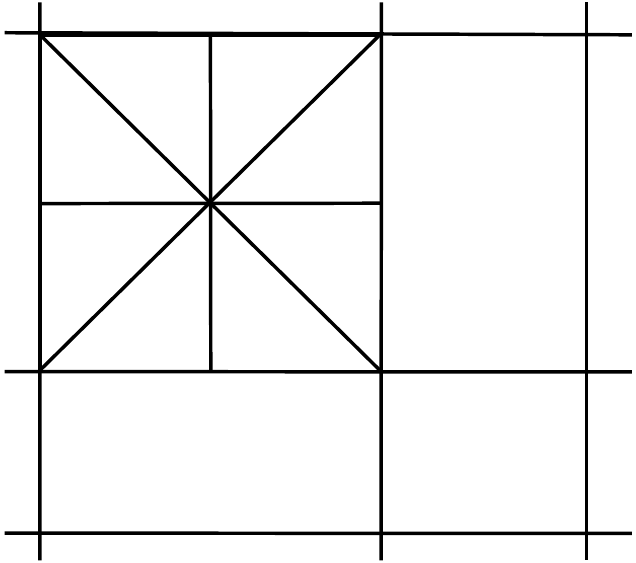}
                \caption{Union}
                \label{fig:Union}
        \end{subfigure}
                ~ 
                \begin{subfigure}[b]{0.23\textwidth}
                \centering
                \includegraphics[width=\textwidth]{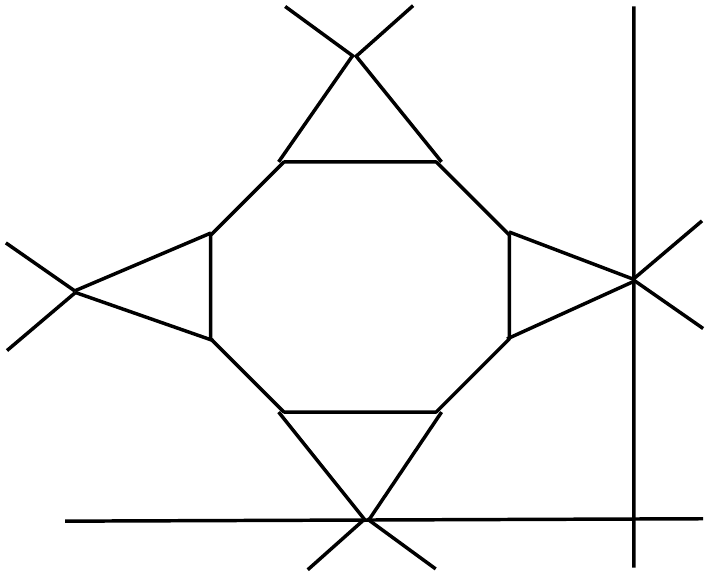}
                \caption{Dual Union}
                \label{fig:DualUnion}
        \end{subfigure}
        \caption{\small The ``reduced square'' and ``union'' lattices with their dual lattices. The dual lattice is obtained from the primal by exchanging faces and vertices with each other.}
        \label{Lattices}
\end{figure}

Our goal is to study the toric code beyond regular self-dual lattices. By considering such lattices, one can break the symmetry between properties of $X$ and $Z$ error-correction. For example, consider the toric code on the reduced square lattice, as shown in Fig. \ref{Lattices}. In this case $\mathcal{L} \neq \bar{\mathcal{L}}$ and hence error-tolerances and lifetimes are no longer symmetric. Intuitively, $e$ anyons are more localised because the primal lattice is less connected leading to a better anyone pairing. On the other hand, $m$ anyons on the dual lattice are more sparse because the dual lattice is more connected allowing anyon pairs to travel far apart. Indeed, the error-tolerance thresholds of the primal and dual lattices are $13\%$ and $8\%$, respectively. Furthermore, the lifetimes of the lattices are $6.8$ and $4.7$, respectively.

\section{Error model and simulations}
In order to determine the lifetime of the memory against thermal noise, it is necessary to simulate the resulting dynamics. In this section we show how to model the classical dynamics of the system. We model the interaction of the system with a thermal environment by coupling each spin to a bath which can introduce spin errors in the initial state $| \psi_0 \rangle $, the ground state of the system \cite{Alicki,James-Beat}. In the limit of weak coupling, the rate for the probabilities $p_{\psi}$ to be in state $|\psi \rangle$ is described by a coupled rate equations 
\begin{equation}
\dot{p}_{\psi} = \sum_{i} \lbrack  \gamma^{in}_{i,\psi} \; p_{z_i(\psi)}(t) - \gamma^{out}_{i,\psi} \; p_{\psi}(t)\rbrack.
\label{RateEquation}
\end{equation}
Here, $p_{\psi}(t)$ is the time-dependent probability to find the system in the state $|\psi \rangle =  \prod_{k \in \psi } Z_k |\psi_0 \rangle$ where $\psi$ is the set of affected spin indices in $| \psi \rangle$. Similarly, $p_{z_i(\psi)}(t)$ is the probability to be in the state $Z_i |\psi\rangle$. Finally, $\gamma^{in} _{i,\psi} = \gamma(-\omega_i,\psi)$ and  $\gamma^{out} _{i,\psi} = \gamma(\omega_i,\psi)$ are the transition rates to arrive or leave the state $|\psi \rangle$, where $\omega = \epsilon_{\psi} - \epsilon_{z_i(\psi)} $ is the energy difference between the states $|\psi \rangle$ and $Z_i |\psi\rangle$.

\begin{figure}[t]

	\begin{subfigure}[b]{0.5\textwidth}
	\centering 
	\includegraphics[width=\textwidth,height=\textwidth]{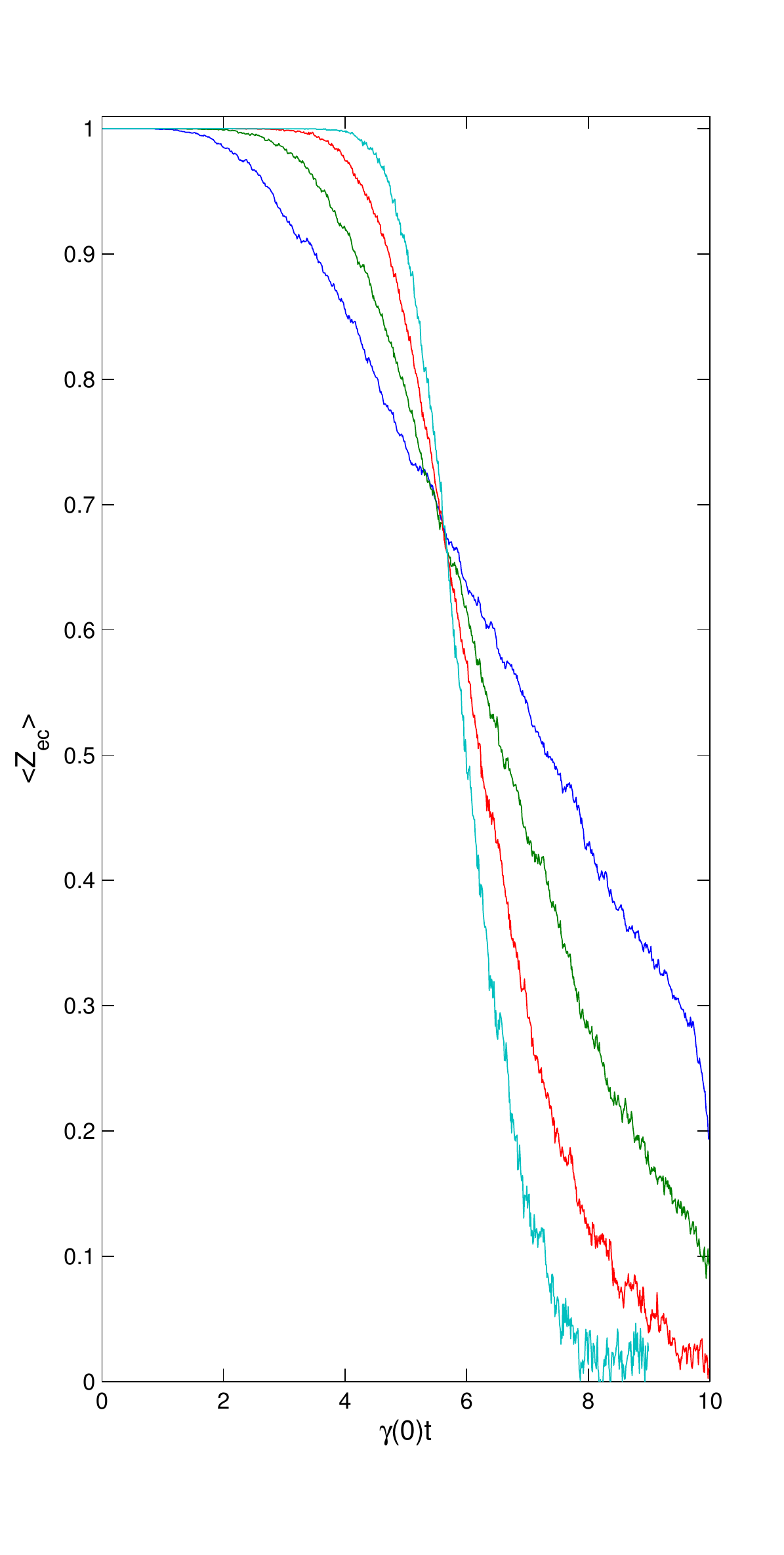}
	\end{subfigure}%
	~
	\begin{subfigure}[b]{0.5\textwidth}
	\centering 
	\includegraphics[width=\textwidth,height=\textwidth]{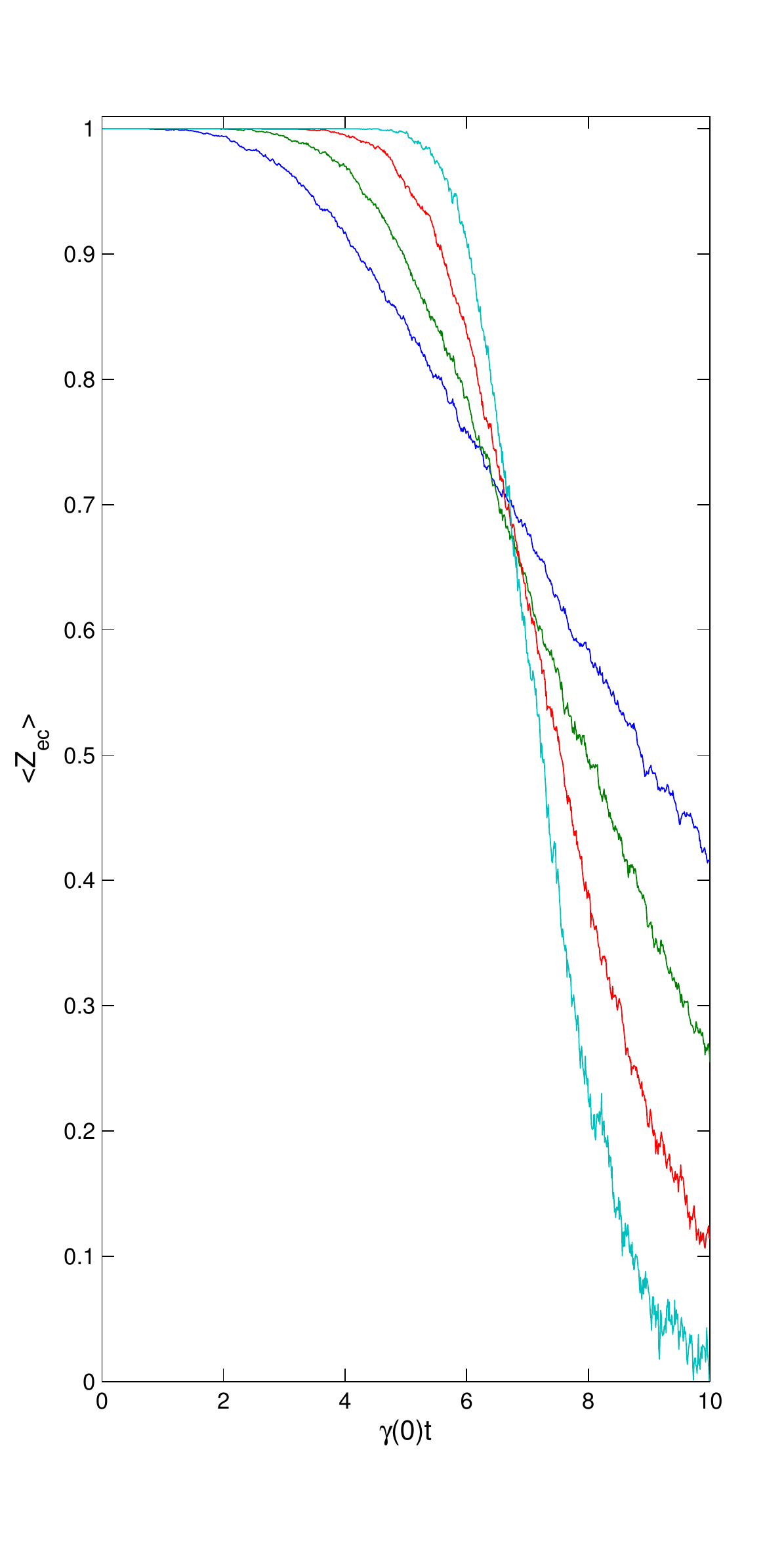}
	\end{subfigure}
	
	\caption{ \small (Colour online.) Decay of the logical Z operator in the non-interacting toric code for the square lattice (left) and reduced square lattice (right) for grid sizes $L=16,32,64,128$ and $L=15,30,60,120$, respectively. The transition becomes sharper as the system size increases. All curves are ensemble averages over $10^{4}$ runs. We  have used $T/J=0.3$, $\gamma(0) = \gamma(2J)$ and applied error correction at the readout stage. }
	
\end{figure}

The rates $\gamma(\omega)$ describe transition probabilities between states with energy difference $w$. A standard expression for $\gamma(\omega)$ can be obtained from a spin-boson model and reads
\begin{equation}
\gamma(\omega) = 2 \kappa_n    \left| \frac{\omega^n}{1-e^{-\beta \omega }} \right| e^{|w|/w_c}
\end{equation}
where $\kappa_n$ is a constant with units $1/ \mbox{energy}^n$ setting the time scale, $\beta = 1/k_BT$, with $T$ being the temperature of the bath and $k_B$ denoting Boltzmann's constant which we will set to one, and $\omega_c$ is the cutoff frequency of the bath \cite{Leggett, Loss}. For simplicity, we assume a large cut off energy $\omega_c \rightarrow 0$. For $n=1$, the bath is called ``Ohmic", whereas for $n \geq 2$ it is called ``super-Ohmic". In this work we will only consider the former case. The relevant rates for our simulation of the toric code are $\gamma(0)$ (rate for an anyon to hop to a free neighbouring site), $\gamma(-2J)$ (rate to create an anyon pair), and $\gamma(2J) = \gamma(-2J) e^{2\beta J}$ (rate to annihilate a pair of adjacent anyons, obtained from the detailed balance property). An important feature of the Ohmic case is that $\gamma(0) \neq 0 $, thus direct hopping to neighbouring sites is always allowed.

It is impossible to solve Eq. (\ref{RateEquation}) analytically for meaningful system sizes, because the number of states $p_{\psi}$ grows exponentially with lattice size. We thus stochastically simulate the system and obtain the quantities of interest. A single iteration of the simulation at time $t$ consists of (i) record the relevant parameters of the system; (ii) calculate the total spin flip rate $ R = \sum_{i} \gamma(\omega_{i,\psi}) $, where $\psi$ is the current state of the system; (iii) draw the time it takes for the next spin-flip to occur from an exponential distribution with rate $R$; (iv) calculate individual spin-flip probabilities $p_i = \gamma(\omega_{i,\psi}) /R$ and flip a spin at random accordingly. After some initially specified time has been reached, we stop and have obtained a single iteration.

We also apply an error correction scheme at the read out stage. The error correction scheme consists of pairing up all detected anyons and then annihilating them by connecting each pair with a string of errors from one partner to another. The pairing is usually chosen such that all anyons are annihilated with the smallest total number of single-spin operators. This is known as the minimal-weight perfect matching and can be found in polynomial time with the help of {\em Blossom} algorithm \cite{Edmonds}. In order to find the true matching with minimal weight, one in principle would need to choose the set of edges that include all connections from every anyon to every other. This forms a complete graph with number of edges that grows quadratically with $n$. Hence, the overall scaling of the matching algorithm becomes formidable for large $n$. Instead, as a good approximation we perform a Delaunay triangulation using Triangle \cite{Shewchuk}, so that searching is performed only on anyons that are close to each other resulting in a number of edges that is linear in the number of anyons.

\section{Degree as a measure of connectivity}
There are a great deal of different lattices on which the toric code may be defined with many distinct characteristics. However, much of their behaviour with respect to the propagation of errors depends on a single property, the connectivity of the lattice \cite{Fujii}. The average degree of a lattice is a simple and effective measure for the connectivity. Lattices with high connectivity will have a large average degree and vice versa. Consider a periodic connected planar lattices with a finite number $K$ such that a unit cell of the lattice can have vertices of various degrees $d_1, d_2, \ldots, d_K$. The degree of a vertex is the number of edges originating from the vertex. The average degree of the lattice is defined as a weighted average 
\begin{equation}
q = \sum^{K}_{i=1} w_id_i,
\label{Lifetime}
\end{equation}
where the weights $w=(w_1, \ldots, w_K)$ are the proportions in which each node appears within a unit cell. Furthermore, by using Euler's formula for planar graphs, one can show that for any periodic lattice 
\begin{equation}
\frac{2}{q} + \frac{2}{\bar{q}} = 1
\label{Euler}
\end{equation}
where $\bar{q}$ is the average degree of the dual lattice $\bar{\mathcal{L}}$. For example, the triangular lattice has an average degree $q=6$, its dual, the hexagonal lattice has an average degree $\bar{q} = 3$. Intuitively, one expects a highly connected lattice (large $q$) to be fragile against $Z$ errors, resulting in a relatively small error-tolerance threshold. On the other hand, by Eq. (\ref{Euler}), its dual lattice will be less connected (small $\bar{q}$) and more robust against $X$ errors, resulting in a relatively bigger error-tolerance threshold.

\begin{figure}[t]
\centering
\includegraphics[width=10cm]{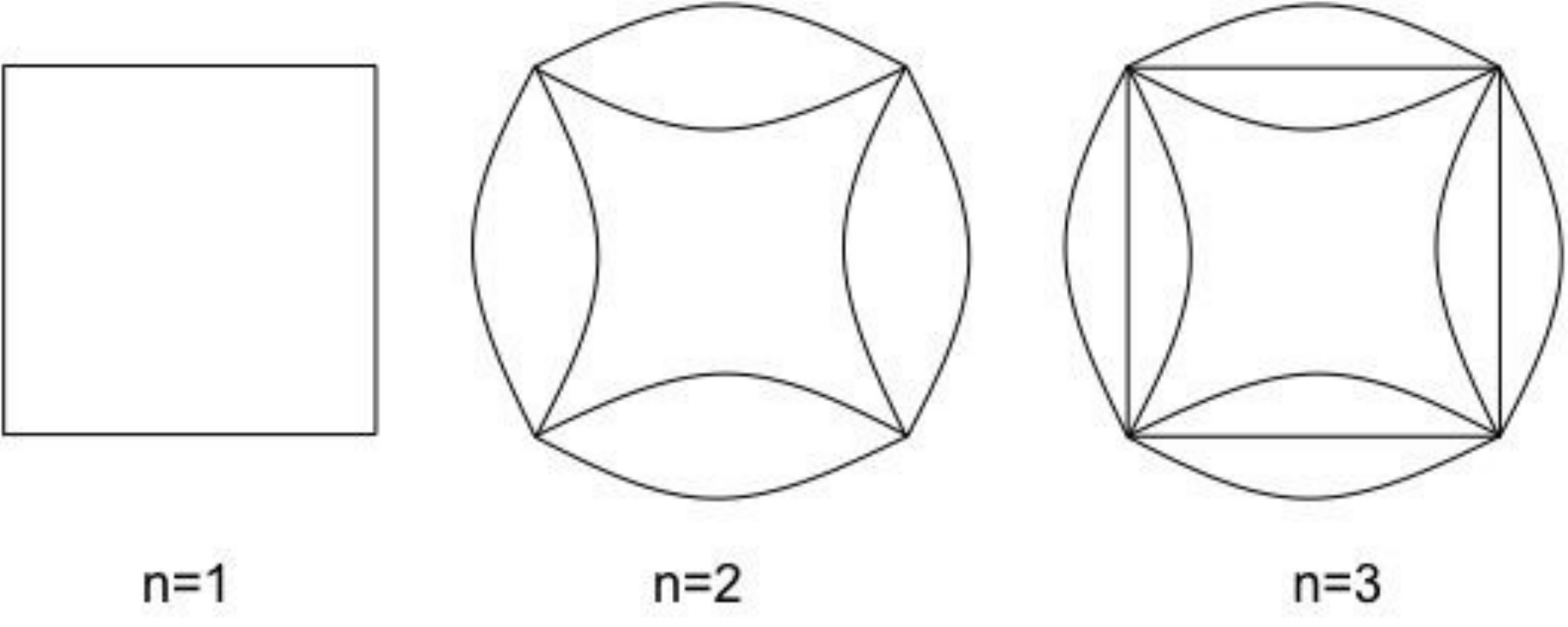}
\caption{\small{Examples of a family of lattices obtained by replacing each edge of the square lattice with two and three sub-edges respectively. In general one replaces each edge of the square lattice with n-sub-edges.}}
\label{fig:Asymp}
\end{figure}

To gain an insight into how the threshold probability varies with average degree, especially in the $q \rightarrow \infty$ limit, we will consider the class of lattices given in Fig. \ref{fig:Asymp}. These are based on the square lattice, but replace each edge with $n$ so-called sub-edges. Each edge therefore contains $n$ spins, rather than just one as before, and each vertex has a degree of $4n$. The effect of a $Z$ error on any of the $n$ spins per edge is identical and indistinguishable to that on any other. Each changes the anyonic occupancy of the two vertices connected by the edge. A net error on the edge therefore only occurs when an odd number of single spin errors are applied, since even numbers of errors cancel each other out. If the probability of a single spin $Z$ error is $p$, the probability of a net error on the edge is therefore
\begin{equation}
\label{P}
P(p) = \frac{1}{2} (1-(1-2p)^{n}).
\end{equation}
Error correction on the $n$-sub-edge lattice is exactly the same as that on the square lattice, except that $P$ is used as the error probability per edge. The threshold probability for the $n$-sub-edge lattice, $p_c^n$, is therefore that for which $P(p_c^n) = p_c$, where $p_c \approx 0.11$ is the normal threshold probability for the square lattice. Rearranging Eq. (\ref{P}), we find,
\begin{eqnarray}
p_c^n  &= \frac{1}{2} (1-(1-2 p_c)^{1/n}) \\
 &=  \frac{1}{2} (1-(1-2 p_c)^{4/q}) \\
  &  =  \frac{1}{2} (1-e^{\frac{4}{q}\ln(1-2p_c)} ) \\ 
   & = (-2/q) \ln(1-2p_c)\approx O(1/q) 
\end{eqnarray}
Here we have used the fact that $n=q/4$. This shows that the threshold probability will decay as $O(1/q)$ as $q \rightarrow \infty$. This decay is consistent with that seen in Fig. \ref{fig:LifeTimes}. For the dual lattices with $\bar{q}_n = 2 + (\bar{q}_0-1/n)$, a similar argument based on the Chernoff bound $p_0 = e^{-2n(p_n-1/2)^2}$ can be used to show that as $1/(\bar{q}-2)\rightarrow \infty$, $p_c \approx 1/2 - O(\sqrt{\bar{q} - 2})$.

\begin{table}[t]
\centering 
\begin{tabular}{l cc  ccccccccc}  
\hline\hline
Lattice & & Average Degree  & $p_c$  & T/J= & 0.28 & 0.29 & 0.3 & 0.31 & 0.32 &
\\ [0.5ex]
\hline 
& &Primal (4.0)    & 0.1  & & 7.3 &  6.4  &  5.7 & 5& 4.6&  \\[-1ex] 
\raisebox{1.5ex}{Square} & \raisebox{1.5ex}{}& Dual (4.0)     & 0.1  & & 7.3 & 6.4 & 5.7 & 5 & 4.6  \\[1ex] \hline
& &Primal (3.56)      & 0.13  & & 9.0&  7.7  &  6.8 & 6.0 & 5.4 \\[-1ex] 
\raisebox{1.5ex}{Reduced Square} & \raisebox{1.5ex}{}& Dual (4.57)     & 0.08  & & 6.1 & 5.3 & 4.7 & 4.1 &3.7 \\[1ex] \hline
&&Primal (4.44)    & 0.09  & &  6.8 &$5.8$&5.0&4.55 & 4.15 \\[-1ex] 
\raisebox{1.5ex}{Union} & \raisebox{1.5ex}{}& Dual (3.63)    & 0.11  & & 8 & 7.2  &  6.5 & 5.65 & 5.0   \\[1ex]\hline
\hline 
\label{tab:Lifetime}
\end{tabular}
\caption{\small{Data for primal and dual, square, reduced square and union lattices. The second column lists  the average degree of each lattice. The third column lists error-tolerance thresholds obtained from simulations. The table also compares the lifetimes of the quantum memories as a function of $T/J$.}} 
\end{table}

\section{Numerical results on the relationship between $d$, $p_c$ and $\tau$}
Now that all our tools have been introduced, we may now begin to a study of how the error tolerance thresholds and thermal lifetimes of lattices depend on their connectivity. We focus on concrete examples of reasonable lattices, i.e. those for which no vertex or plaquette operator need act on an nonphysically high number of spins.

\begin{figure}[t]

	\begin{subfigure}[b]{0.5\textwidth}
	\centering 
	\includegraphics[width=\textwidth,height=\textwidth]{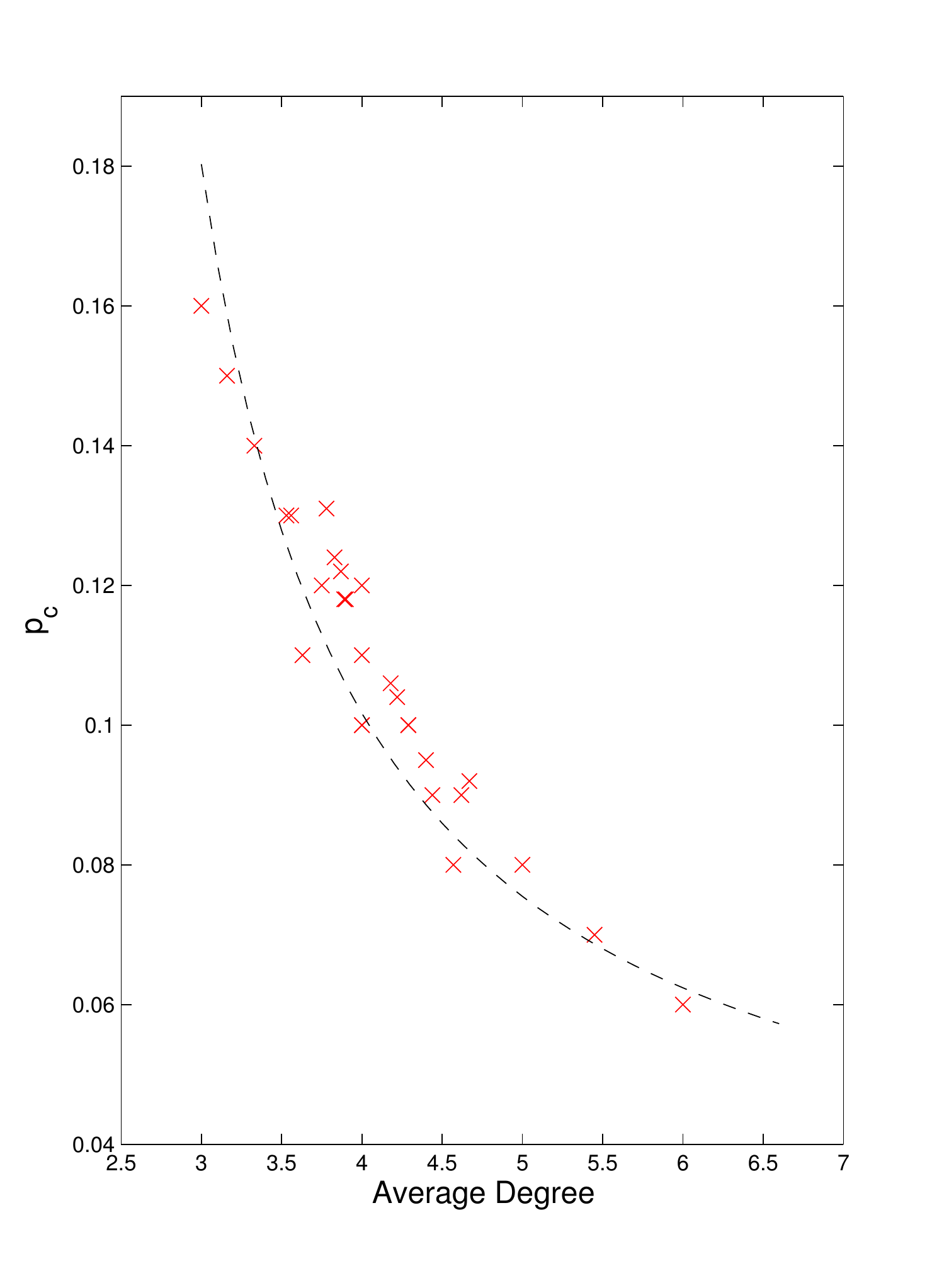}
	\end{subfigure}%
	\;\;\;\;
	\begin{subfigure}[b]{0.5\textwidth}
	\centering 
	\includegraphics[width=\textwidth,height=\textwidth]{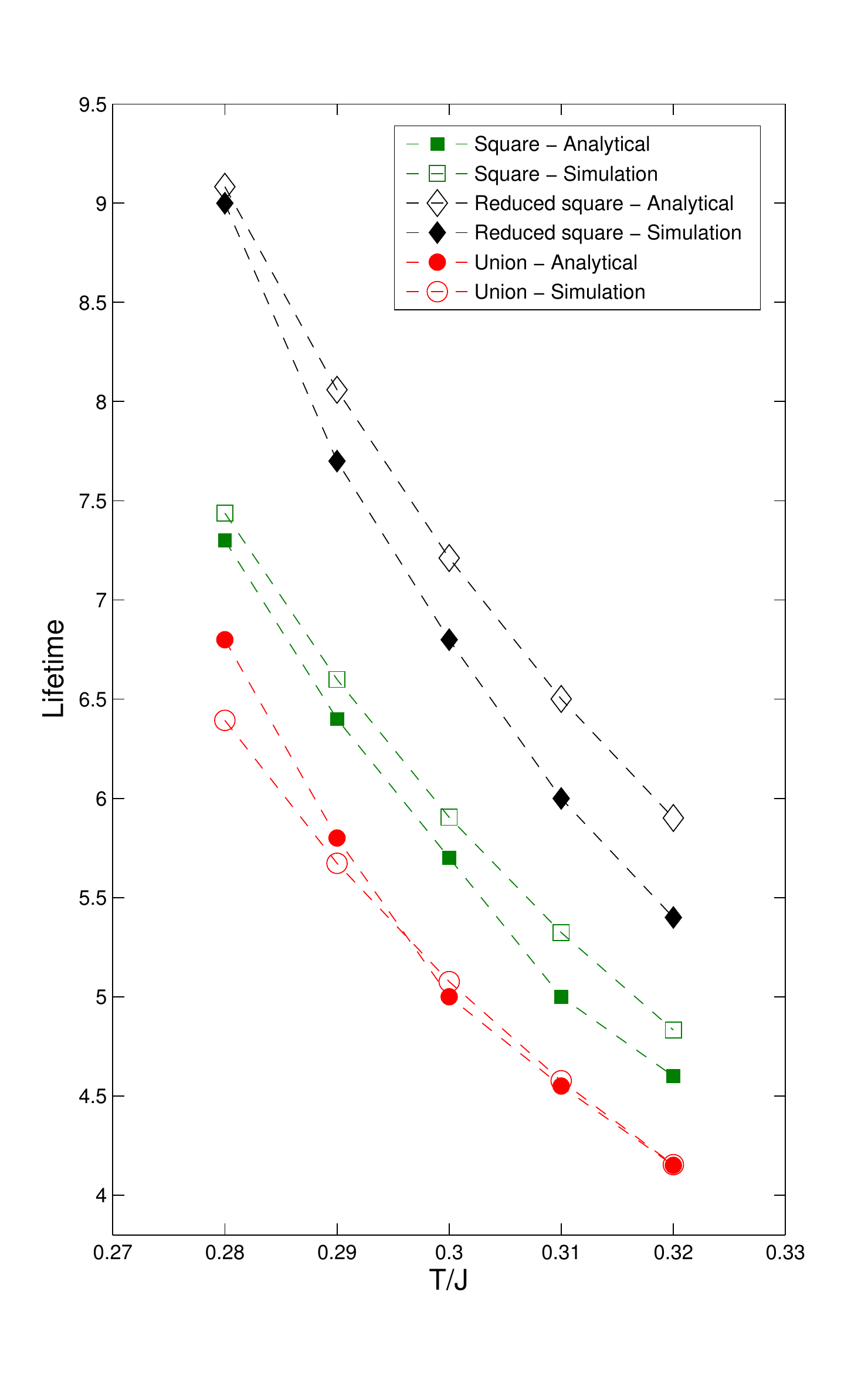}
		\end{subfigure}%
		\caption{\small(Colour online.) {\em Left:} Simulation data of error-tolerance probability thresholds for lattices with various average degrees. We have focused on realistic lattice geometries with average degree $ 3 \leq q \leq 6$.
{\em Right:} Analytical data obtained from Eq. (\ref{LifetimeDegree}) versus simulation data for the lifetime of the quantum memory for various values of $T/J$ and for various lattice patterns. This comparison reveals that the Eq. (\ref{LifetimeDegree}) provides a good description of the system's lifetime.} 
	 \label{fig:LifeTimes}
\end{figure}

Fig. \ref{fig:LifeTimes} shows the error-tolerances for various lattices as a function of the degree of the lattice. For a square lattice where $\mathcal{L} = \bar{\mathcal{L}}$, the error-tolerances of the primal and dual lattices is symmetric. One can break this by considering non-self-dual lattices. From Eq. (\ref{Euler}), we know that for such lattices, there is an asymmetry in the average degrees of the primal and dual lattices. This in turn results in an asymmetry in the error-tolerance thresholds of the primal and dual lattices. The data in Fig. \ref{fig:LifeTimes} confirm our earlier intuition that lattices with small average degrees (i.e. less connectivity) will have better localisation properties and consequently high error-tolerance thresholds. On the other hand, lattices with large average degrees (i.e. low connectivity) will have less localisation properties and consequently small error-tolerance thresholds.

The data in Table (1) show the lifetime of the memory for several lattices at different values of $T/J$. Clearly, the higher the temperature $T$, the smaller the lifetime. Furthermore, for any lattice with $\mathcal{L} \neq \bar{\mathcal{L}}$, the lifetimes of the $\mathcal{L}$ and $\bar{\mathcal{L}}$ are not the same. Consequently any formula for the lifetime must take into account the structure of the underlying lattice, as well as the coupling constant $J$ and the temperature of the system $T$. This because for any non-self-dual lattice the lifetime must exhibits an asymmetry and varies with $J$ and $T$. Let $N$ be the number of diffusing anyons present in the system, the fraction $p$ of spins affected by $Z$ error after a time $t$ is estimated as, 
 \begin{equation}
 p = \frac{\varepsilon \,t N}{L^2},
 \end{equation}
where $\varepsilon$ is some constant that may or may not depend on the lattice structure. The memory fails when $p$ is larger than some critical value $p_c$.  This gives a lifetime $\tau$ for the memory
\begin{equation}
\tau = \frac{p_c}{\varepsilon}(e^{J/T} + 1),
\label{LifetimeC}
\end{equation}
where we have replaced the density of anyons by the Fermi-Dirac equilibrium distribution $N / L^2 = (e^{J/T} + 1)^{-1} $. It turns out that the constant $\varepsilon$ is actually independent of the lattice size and structure and that $\varepsilon \approx 1/2$. Therefore 
\begin{equation}
\tau \approx {2p_c}(e^{J/T} + 1).
\label{Lifetime}
\end{equation}
This formula gives an excellent approximation for the memory lifetime when the error-tolerance threshold is known. Furthermore, one can rewrite this formula by expressing $p_c$ it in terms of the average degree $q$ 
\begin{equation}
\tau \approx \frac{2 (\mu q + \nu)}{q-2}(e^{J/T} + 1).
\label{LifetimeDegree}
\end{equation}
where $\mu = 0.0231$ and $\nu =0.111$ are the fitting parameters obtained from Fig. \ref{fig:LifeTimes}. Thus, by simply  knowing the average degree of a lattice at a given $T/J$, one can estimate the lifetime of the quantum memory. Fig. \ref{fig:LifeTimes} compares the lifetimes of the memory obtained using Eq. (\ref{Lifetime}) with those obtained by direct simulation. Despite the approximation used in deriving Eq. (\ref{Lifetime}), the lifetimes obtained from the two approaches are very close.

An important conclusion we draw from the above analysis is that for a given value of $T/J$, one can not hope to improve the lifetimes of both the primal and dual lattice, relative to the square lattices, solely by changing the lattice geometry. However, under a given temperature $T$, if one consider physical systems with biased couplings $J$ for the vertex and the plaquette interactions, then one can find an optimal lattice structure that improves the lifetimes of both the primal and dual lattices relative to the square lattice under the same temperature $T$.

\section{Josephson-junction arrays}
\begin{figure}[t]
\centering
\includegraphics[width=8cm,height=0.7\textwidth]{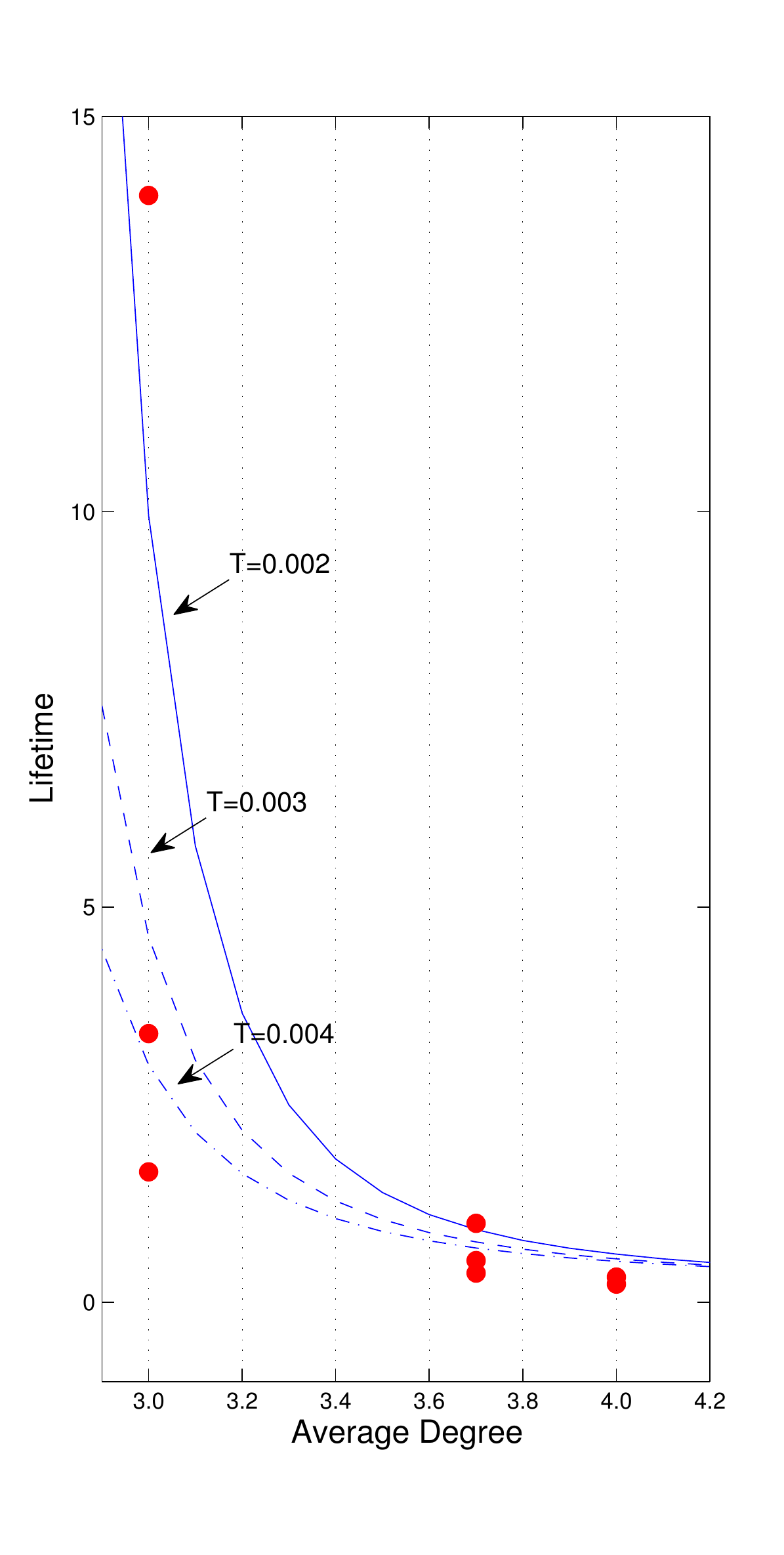}
\caption{The coherence time (minimum of $\tau_p$ and $\tau_d$) of the quantum memory obtained from Eqs. (\ref{CoherenceP}) and (\ref{CoherenceD}) as a function of the average degree for $x=1.01$. From the top, each curve represent the coherence time at temperature $T=0.002, 0.003, 0.004$. The circles are the values of the coherence time obtained from direct simulations.}
\label{fig:LifetimeJj}
\end{figure}

In this section we demonstrate that for systems with biased couplings, it is possible to improve the coherence time (the minimum of $\tau_p$ and $\tau_d$)  of the quantum memory relative to the square lattice by optimising lattice geometry. We demonstrate this by considering a Josephson-junction implementation of the quantum memory \cite{Doucot}. The relations that relate the charging and the Josephson energies, $E_C$ and $E_J$, to the couplings are given by
\begin{eqnarray}
J_{v} &= E_{J}^{3/4} E_C^{1/4} \exp(-q.\alpha \; \sqrt{E_J/E_C}), \\
J_{f} &= \frac{E_J}{\bar{q}},
\end{eqnarray}
where the subscripts $v$ and $f$ refer to the vertex and face operators respectively, $\alpha=1.61$ is a constant and $q$ and $\bar{q}$ are related by Eq. (\ref{Euler}) \cite{Doucot}. Let $x$ be the ratio $x = E_J/E_c$, for $x>1$ the system is said to be in the toric code regime. Using this we can simplify the above relations
\begin{eqnarray}
J_{v}/E_C &= x^{3/4} \exp(-q.\alpha \; \sqrt{x}), \label{JvCoupling} \\
J_{f}/E_C &= \frac{x}{\bar{q}}.
\label{JfCoupling}
\end{eqnarray}
These formulas relate the couplings to the structure of the lattice. We can substitute Eqs. (\ref{JvCoupling}), (\ref{JfCoupling}) in the lifetime formula to obtain formulas for the lifetimes of the primal and dual lattices in terms of $q$ and $x$
\begin{eqnarray}
\tau_p & \approx \frac{2(\mu q + \nu)}{q-2} \exp( \frac{x^{3/4}}{T} e^{-(1.61q\sqrt{x})}), \label{CoherenceP}\\
\tau_d & \approx \frac{2(\mu \bar{q} + \nu)}{\bar{q}-2}  \exp(\frac{x}{T} \bar{q}). \label{CoherenceD}
\end{eqnarray}
Thus, for a fixed $x$, by optimising the degree of the lattice $q$ one hopes to be able to improve the coherence time of the memory relative to the square lattice. For example, using these formulas, for $x=1.01$, we calculated the lifetimes  as a function of the degree for different temperatures, Fig. \ref{fig:LifetimeJj}. It is evident that along a given curve, say $T=0.003$, one can achieve a better coherence time for a Josephson-junction quantum memory by considering, say, a hexagonal lattice ($q=3$) rather than a square lattice $(q=4)$.
\section{Conclusion}
We have studied the effect of deforming lattice geometry on the error-tolerance and lifetime of the quantum memory. We have considered the toric code on non-regular lattices obtained by removing or adding edges to the square lattice. Unlike square lattice, these lattices are non-self-dual, hence, the connectivity and the error-tolerance properties of the primal and the dual lattices are no longer the same. We have shown that the average degree of a lattice is a good measure for its connectivity and by considering families of lattices we have shown that the two quantitates are closely related. Lattices with small average degree will be more robust against errors and vice versa. We have also studied the lifetime of these lattices and provided a formula for calculating the lifetime of the memory at a given temperature by using the average degree. Finally, we have shown that for systems with biased couplings such as Josephson -junction arrays, it is possible to improve the coherence time of the memory relative to the square lattice by deforming the lattice geometry.

\flushleft{{\bf Acknowledgements}}

The authors would like to thank Beat R\"othlisberger  for providing some of the data used in Fig. \ref{fig:LifeTimes}. This work was supported by EPSRC, the Swiss NF, NCCR Nano and NCCR QSIT.

\end{document}